\documentclass{llncs} %
\usepackage{float}
\usepackage{graphicx}
\usepackage{wrapfig}

\usepackage[labelfont=bf]{caption}
\usepackage[colorlinks=true,linkcolor=blue,citecolor=blue]{hyperref}

\usepackage{graphicx}
\usepackage{color,soul}
\usepackage{amsmath} % Mathe
\usepackage{mathtools} % Mathe
\usepackage{amsfonts} % Mathesymbole
\usepackage{amssymb}
 
\usepackage{array} 
\usepackage{booktabs}
\usepackage{rotating}
\usepackage{multirow}
\usepackage{multicol}
\usepackage{lscape}
\usepackage{subfig}
\usepackage{caption}

\usepackage{algorithm}
\usepackage[noend]{algpseudocode}

\algrenewcommand\algorithmicforall{\textbf{foreach}}
\algrenewcommand\algorithmicindent{.8em}
\usepackage[export]{adjustbox}

\usepackage{tikz,xcolor,hyperref}

\definecolor{lime}{HTML}{A6CE39}
\DeclareRobustCommand{\orcidicon}{
	\begin{tikzpicture}
	\draw[lime, fill=lime] (0,0) 
	circle [radius=0.16] 
	node[white] {{\fontfamily{qag}\selectfont \tiny ID}};
	\draw[white, fill=white] (-0.0625,0.095) 
	circle [radius=0.007];
	\end{tikzpicture}
	\hspace{-2mm}
}

\foreach \x in {A, ..., Z}{\expandafter\xdef\csname orcid\x\endcsname{\noexpand\href{https://orcid.org/\csname orcidauthor\x\endcsname}
			{\noexpand\orcidicon}}
}
			
\definecolor{darkgreen}{rgb}{0.53, 0.66, 0.42}

\begin{document}

\title{Meta-RegGNN: Predicting Verbal and Full-Scale Intelligence Scores using Graph Neural Networks and Meta-Learning}

\titlerunning{Meta-RegGNN: Predicting Verbal and Full-Scale Intelligence Scores}  % abbreviated title (for running head)

\author{Imen Jegham\orcidC{}\index{Jegham, Imen}\inst{1,2,3} \and Islem Rekik\orcidA{} \index{Rekik, Islem}\inst{2}\thanks{ {corresponding author: \url{irekik@itu.edu.tr}.}} }

\authorrunning{I Jegham et al.}

\institute{$^{1}$ Universit\'e de Sousse, Ecole Nationale d'Ing\'enieurs de Sousse, LATIS- Laboratory of Advanced Technology and Intelligent Systems, 4023, Sousse, Tunisie;\\
$^{2}$ Horizon School of Digital Technologies, 4023, Sousse, Tunisie; \\
$^{3}$ BASIRA Lab, Faculty of Computer and Informatics Engineering, Istanbul Technical University, Istanbul, Turkey (\url{http://basira-lab.com/})}

\maketitle              % typeset the title of the contribution 

\begin{abstract}
Decrypting intelligence from the human brain construct is vital in the detection of particular neurological disorders. Recently, functional brain connectomes have been used successfully to predict behavioral scores. However, state-of-the-art methods, on one hand, neglect the topological properties of the connectomes and, on the other hand, fail to solve the high inter-subject brain heterogeneity. To address these limitations, we propose a novel regression graph neural network through meta-learning namely Meta-RegGNN for predicting behavioral scores from brain connectomes. The parameters of our proposed regression GNN are explicitly trained so that a small number of gradient steps combined with a small training data amount produces a good generalization to unseen brain connectomes. Our results on verbal and full-scale intelligence quotient (IQ) prediction outperform existing methods in both neurotypical and autism spectrum disorder cohorts. Furthermore, we show that our proposed approach ensures generalizability, particularly for autistic subjects. Our  Meta-RegGNN source code is available at \url{https://github.com/basiralab/Meta-RegGNN}.
\end{abstract}


%% ***************************************************************************** %%
\section{Introduction}
%% ***************************************************************************** %%

Autism, or Autism Spectrum Disorder (ASD), is a neurodevelopmental disorder that affects how a person feels, thinks, interacts with others, and encounters their environment. Research has shown that subjects with ASD have higher rates of health issues throughout childhood, adolescence, and adulthood and this can lead to a high risk of early mortality. ASD diagnosis remains a challenging task due to the wide range in the severity of its symptoms and the lack of a pathophysiological marker
\cite{doi:10.1177/11786329221078803,hodges2020autism}. Recently, machine learning techniques have become a primary route for computer-aided diagnosis, and have been broadly used to analyze autism disorders \cite{rahman2020review,xu2021brain,hyde2019applications}. Intelligence, in particular, is a key aspect of ASD. State-of-the-art methods successfully used functional brain connectomes to predict cognitive measures such as Intelligence Quotient (IQ) scores in both disordered and healthy cohorts \cite{dryburgh2020predicting,hanik2022predicting,yamin2020geodesic}. Indeed, functional brain connectomes describe the brain network structure and are derived from resting-state magnetic resonance imaging (MRI). They are modeled as graphs whose nodes depict anatomical regions of interest (ROIs) and  edges represent the correlations in activity between ROI pairs \cite{liu2021graph}. 

To improve generalizability across contexts and populations, Shen et al. \cite{shen2017using} developed a data-driven protocol for Connectome-based Predictive Modeling (CPM) of brain-behavior relationships by training linear regression model using cross-validation. To ameliorate the obtained results, Dryburgh et al. \cite{dryburgh2020predicting} studied how neural correlates of intelligence scores are altered by atypical neurodevelopmental disorders by performing their analysis in both Neuro Typical (NT) subjects and subjects with ASD. For that, they adopted CPM and evaluated negative and positive correlations of brain regions separately. However, these methods flatten the brain connectome matrix though vectorization which neglects the graph structure of the connectomes. Thus, the local and global topological properties of the connectomes that are rich of information are not exploited.

To overcome this issue, Graph Neural Networks (GNNs) have been proposed. They can handle complex graph data and have proven their exclusive ability in learning in non-Euclidean spaces including graphs with complex topologies and a wide range of graphs \cite{he2020deep}. GNN is firstly proposed in 2005 \cite{gori2005new} to be then elaborated on in detail \cite{scarselli2008graph}. GNNs are a class of deep learning techniques with graph convolutional layers that outperform existing methods in a large range of computer vision applications \cite{bessadok2021graph}. Recently, they have received large attention thanks to their exclusive ability in effectively modeling the correlation between samples. They provide an efficient solution to integrate diverse information. However, a lack of works that explored GNN for the prediction of cognitive scores has been noticed. Hanik et al. \cite{hanik2022predicting} was the first to propose a GNN architecture, called RegGNN, specialized in regressing brain connectomes to a cognitive score to predict. To better improve the performance of GNN, they also proposed a learning-based sample selection method that selects training samples with the highest predictive power. However, existing GNN-based models present a major drawback which is the lack of flexibility which means that the model fails to be used for independent testing \cite{song2021auto}.

As a key challenge for the cognitive score prediction is high heterogeneity across individual brains, standard learning approaches fail when applied in different conditions than used for training. To decrease this covariate shift that drastically affects the usefulness of machine learning models and improve the generalizability of proposed methods, meta-learning approaches have been proposed and achieved a tremendous success in recent years \cite{wang2021meta}.
The basic idea of meta-learning or learn to learn is to gradually enhance the performance of a model by learning multiple different tasks. It is similar to transfer learning \cite{torrey2010transfer}. In transfer learning, model parameters are learned  after being trained with lots of data and then fine-tuned to obtain good parameters, while in meta-learning, good model parameters that are sensitive to small changes and give large improvement on loss function for a particular task are learned. Meta-learning aims to rapidly learn a new task from a small amount of new data, and the model is trained by the meta-learner to be able to learn on several existing tasks \cite{bai2021important}. There are different meta-learning approaches including one-shot learning with memory augmented neural networks \cite{santoro2016meta}, optimization as a model for few-shot learning \cite{Ravi2017OptimizationAA} and Model Agnostic Meta-Learning (MAML) \cite{finn2017model}. The latter may be directly applied to any learning model that is trained with a gradient descent procedure. With minimal modification, it can simply manage several architectures and multiple problem settings, including policy gradient reinforcement learning, classification and regression. However, despite their important role to ensure generalizability and solve data fracture problem, this method has not been previously employed in predicting cognitive scores.

In this paper, we introduce \emph{the first regression GNN network through meta-learning}, namely Meta-RegGNN that regresses functional brain connectomes to predict cognitive scores. Our Meta-RegGNN network on one hand properly includes the graph structure of functional brain connectomes and effectively models the correlation between them, and on the other hand, thanks to meta-learning, makes the regression GNN model more flexible while decreasing the impact of the high brain variability and domain fracture issues.

The main contributions of our method can be summarized as follows:
\begin{enumerate}
\item We introduce a novel meta-learning regression graph neural network that shows an exclusive ability in modeling the correlation between data and incorporates global and local topological properties of the functional brain connectomes to predict behavioral scores.

\item We present the first work on meta-learning for regression graph neural networks rooted in inductive learning and which boosts the prediction performance by decreasing the effect of sample heterogeneity. This network shows a good trade-off between flexibility and performance and can be used in other application fields suffering from high intra-class variability issues.

\item We illustrate a pipeline, consisting of Meta-RegGNN, which outperforms state-of-the-art models in predicting Verbal Intelligence Quotient (VIQ) and Full-scale Intelligence Quotient (FIQ) from functional brain connectomes in neurotypical and autism spectrum disorder cohorts.
\end{enumerate}

% %% ***************************************************************************** %%
\section{Methodology}
% %% ***************************************************************************** %%

\begin{figure}[ht!]
\centering
\includegraphics[width=1.2\textwidth]{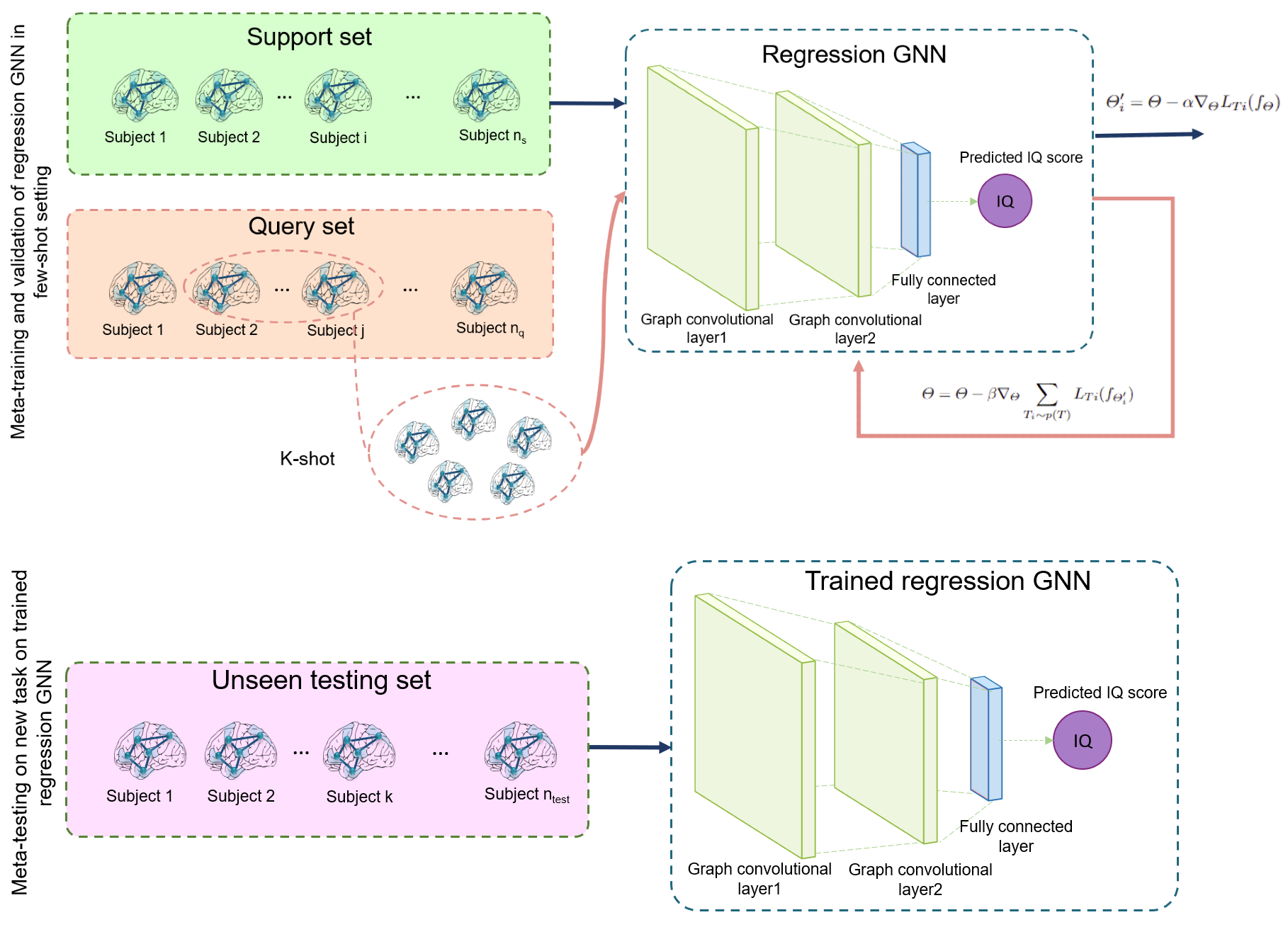}
\caption{Illustration of the proposed meta-training and validation of regression GNN in a few-shot setting.} 
\label{fig:1}
\end{figure}

In this section, we detail the architecture and the algorithm of our proposed.  \textbf{Fig.}~\ref{fig:1} shows the layout of the overall process of Meta-RegGNN. In our proposed approach, meta-training is implemented as episodic tasks on support and query sets. A few-shot learning framework is used for the query set. The goal of this few-shot regression is to predict the behavioral scores from only a few samples after training on many samples with similar statistical properties. During the meta-testing, predicted behavioral scores are obtained using unseen samples that are provided with the optimized weights obtained from the meta-learning stage.

$\bullet$ \textbf{Problem statement.} We consider a regression GNN model, denoted $f$, that maps brain graphs $g$ to behavioral scores $s$. During meta-learning, the regression GNN model is trained to be able to adapt to a large number of tasks. We present a generic notion of a learning task below. Each task
$T = \{L(g_1, s_1, . . . , g_H, s_H), q(g_1), q(g_{t+1}|g_t, s_t),H\}$ consists of a loss function $L$, a distribution over initial observations $q(g_1)$, a transition distribution $q(g_{t+1}|g_t, s_t)$ and an episode length $H$ (in our case, we can define $H=1$ and drop the time-step $t$ on $x_t$, as the model is used for supervised learning and accepts one input and gives one output). For regression, the loss function is defined as follows:
\begin{equation}
L_{Ti}=  \sum_{g^{(j)},s^{(j)}\sim T_i} \|f(g^{(j)})-s^{(j)}\|^2_2,
\label{loss}
\end{equation}

where $g^{(j)},s^{(j)}$ represent respectively the input and output sampled from task $T_i$. In our model agnostic meta-learning scenario, we define a distribution over tasks $p(T)$ that we want our regression GNN model to adapt to. In the K-shot learning setting, the regression GNN is trained to learn a new task $T_i$ from $p(T)$ from only K samples drawn from $q_i$ and the feedback $L_{Ti}$ produced by $T_i$. At the end of meta-training, new tasks are sampled from $p(T)$, and meta-performance is measured by the model’s performance after learning from K samples. 

$\bullet$ \textbf{Meta-RegGNN algorithm.} The aim of our Meta-RegGNN is to prepare our regression GNN model for fast adaptation. Thus, the GNN might learn internal features of functinal brain connectomes that are relevant to all tasks in $p(T)$. For that, we first find the RegGNN model parameters that are responsive to modifications in the given task, so that small modifications in the parameters produce large improvements on the loss function of any task from $p(T)$. Let us consider our regression GNN model represented by a parametrized function $f_ \Theta$ with parameters  $\Theta$. The latter is updated to $\Theta'$ when adapting to a new task $T_i$. The updated $\Theta$ is defined as:

\begin{equation}
\Theta'_i= \Theta- \gamma \nabla_\Theta L_{Ti}(f_\Theta),
\label{theta}
\end{equation}

where $\gamma$ represents the step size hyperparameter. The meta-optimization is achieved over the regression GNN model parameters $\Theta$, while the objective is calculated using the updated regression GNN model parameters $\Theta$'. Indeed, our Meta-RegGNN aims to optimize the model parameters so that one or a small number of gradient steps on a new task generate effective behavior. 

The meta-optimization through tasks is conceived in order to update the regression GNN model parameters $\Theta$ as follows:

\begin{equation}
\Theta= \Theta- \eta \nabla_\Theta \sum _{T_i\sim p(T)} L_{Ti}(f_{\Theta'_i})
\label{theta2}
\end{equation}
where $\eta$ presents the meta-step size. The meta-training algorithm is outlined in Algorithm 1.
\begin{algorithm}
\caption{Meta-training regression GNN  algorithm.}\label{algo}
\begin{algorithmic}[1]
\Require {p(T)= Distribution over tasks}
\Require {$\gamma$, $\eta$: Step size hyperparameters}
\State Initialize $\Theta$ randomly
\While {not done}
  \State  Sample tasks batch $T_i\sim p(T)$ 
  \ForAll{$Ti$}
\State Randomly choose $k$ samples $D=\{g^{(i)},s^{(i)}\}$ from $Ti$
\State Evaluate $\nabla_\Theta L_{Ti}(f_\Theta)$ with respect to $k$ using $D$ and $L_{Ti}$ in Equation \ref{loss}
\State Compute adapted parameters $\Theta'_i$ according to Equation \ref{theta}
  \EndFor

  \State Update $\Theta$ according to Equation \ref{theta2} using $L_{Ti}$ in Equation \ref{loss}
\EndWhile
\State \textbf{end}
\end{algorithmic}
\end{algorithm}

$\bullet$ \textbf{Regression GNN.} To properly take into account the graph structure of the brain connectomes and effectively model the correlation between data samples, we used a regression GNN network that consists of two graph convolution layers and a fully connected layer (\textbf{Fig.}~\ref{fig:1}). Given a correlation matrix of a connectome $C$ is symmetric, that can have zero or positive eigenvalues, we may simply regularize it to be symmetric positive definite according to:
\begin{equation}
I'= C+\mu I,
\label{regularize}
\end{equation}

where $I$ represents the identity matrix and $\mu > 0$ \cite{wong2018riemannian}. In fact, since positive correlations have been demonstrated to be more important in analyzing brain networks \cite{fornito2016fundamentals},  all negative eigenvalues are set to zero to train our regression GNN \cite{hanik2022predicting}. Thus, regression GNN receives the regularized positive adjacency matrix $I'$ of a connectome and predicts the corresponding behavioral scores using graph convolutions. This reduces the size of the brain connectomes and learns an embedding for the brain connectomes. After the first graph convolution operation, we add a dropout layer for regularization. Finally, the obtained embedding goes through a fully connected layer which produces a scalar output (IQ scores). %We note that in this work, we use RegGNN \cite{hanik2022predicting} without the sample selection component.

% %% ***************************************************************************** %%
\section{Experimental results and discussion}
% %% ***************************************************************************** %%
% %% ---------------------------

\textbf{Evaluation dataset.} 
To highlight the utility of our proposed Meta-RegGNN, we evaluated our method on subjects drawn from the Autism Brain Imaging Data Exchange (ABIDE) preprocessed dataset \cite{craddock2013neuro}.  The preprocessed datasets are available online \footnote{http://preprocessed-connectomes-project.org/abide/}. They contain two cohorts: ASD and NT. The ASD cohort comprises  202 patients (with mean age = (15.4 $\pm$ 3.8)), while the NT cohort includes 226 subjects (with mean age = (15 $\pm$ 3.6)). VIQ and FIQ scores in the ASD cohort have means 106.102 $\pm$ 15.045 and 103.005 $\pm$ 16.874 whereas VIQ and FIQ scores in the NT cohort have means 111.573 $\pm$ 12.056 and 112.787 $\pm$ 12.018, respectively. The connectomes of the brain were derived from resting-state fMRI using the parcellation from \cite{tzourio2002automated} into 116 ROIs.

\textbf{Parameter settings.} To evaluate the generazabilty and the effectiveness of our Meta-RegGNN, we used 3-fold cross-validation on ASD and NT cohorts for VIQ and FIQ prediction. Based on empirical observations, we trained our proposed method for 300 epochs with a weight decay at 0.0005 and a learning rate of 0.001. The dropout rate was set to 0.2.  For the meta-training, we used one gradient update with K=5 shots with a step size $\gamma=10 ^{-7}$ and employed Adam optimizer as meta-optimizer \cite{kingma2014adam}. For all methods, we state the Mean Absolute Error (MAE) and the Root Mean Squared Error (RMSE).

\begin{figure}[ht!]
\centering
\includegraphics[width=\textwidth]{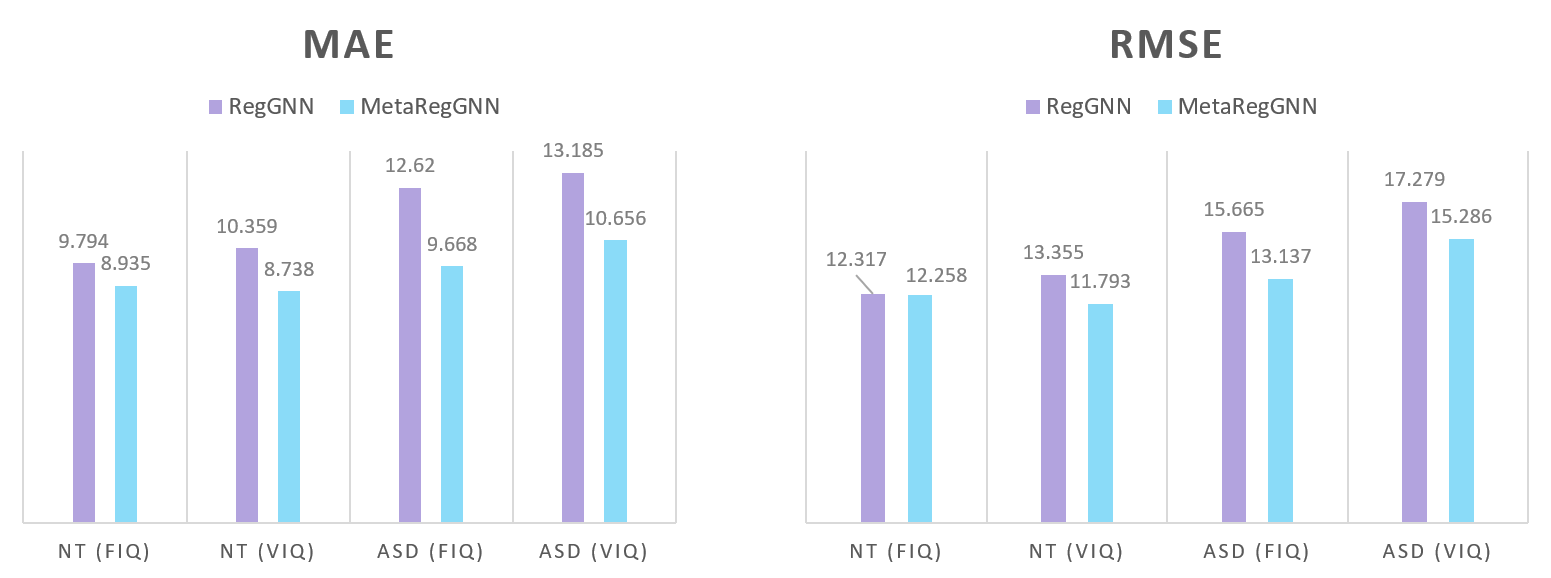}
\caption{Cognitive scores prediction results using different evaluation metrics on the NT and ASD cohorts.} 
\label{fig:2}
\end{figure}

\textbf{Evaluation and comparison method.} To benchmark our method, we chose the first and unique deep learning method proposed in the literature that uses GNN to predict cognitive scores \cite{hanik2022predicting} without the proposed sample selection step. The results for the ASD and NT cohorts for FIQ and VIQ are shown in \textbf{Fig.}~\ref{fig:2}. These results present the average of more than 40 random repetitions of our 3-fold cross-validation.

Compared to the NT cohort, the ASD cohort achieved the worst results across all methods. The difficulty of predicting behavioral scores in the ASD cohort may be explained by the high inter-subject heterogeneity \cite{tordjman2017reframing}. A general improvement by our Meta-RegGNN is noticed in all learning tasks. Our method dealt with the correlation of functional brain connectomes and combined the prior knowledge with automatically learned similarity. Therefore, a high improvement in the ASD cohort is recorded that can be explained by the generalizability improvement. Even with the repeated randomized runs, our Meta-RegGNN displayed the lowest prediction error across both cohorts and metrics, which indicates the stability of our model under data distribution shifts. The best results in terms of MAE and RMSE are noted in the NT cohort which may be explained by the similarity between neurotypical brains. 

Compared with previous studies on predicting behavioral scores, our model achieved a good trade-off between flexibility and performance requiring fewer samples for training. Moreover, it can deal with test samples that are different from those of the training samples (brains diagnosed with Alzheimer’s Disease for example). Despite its multiple advantages, this prime work needs to be further validated on other datasets and different brain connectivity classes.

% %% ***************************************************************************** %%
\section{Conclusion}
% %% ***************************************************************************** %%
In this paper, we proposed the first GNN for regression through meta-learning namely Meta-RegGNN, for behavioral score prediction from brain connectomes. Our network nicely provides an efficient solution which handles the the topological properties of functional brain connectomes. Furthermore, it ensures model flexibility and enables inductive learning, thereby enhancing the model generalizability to unseen data. Our key contributions consist in designing a graph neural network for regression that predicts behavioral scores and training our GNN via model agnostic meta-learning. Our proposed method outperforms state-of-the-art methods in terms of prediction results. In our future work, we will investigate the explainability aspect of our Meta-RegGNN in order to identify connectivity biomarkers distinguishing between  typical and atypical brain states.

% %% ***************************************************************************** %%
\section{Supplementary material}
% %% ***************************************************************************** %%

We provide three supplementary items for reproducible and open science:

\begin{enumerate}
	\item A 7-mn YouTube video explaining how our framework works on BASIRA YouTube channel at \url{https://youtu.be/MS6oXzr1NNg}.
	\item Meta-RegGNN code in Python on GitHub at \url{https://github.com/basiralab/Meta-RegGNN}. 
	\item A GitHub video code demo on BASIRA YouTube channel at \url{https://youtu.be/Fl7DXVEWA8g}. 
\end{enumerate}

% %% ***************************************************************************** %%
\section{Acknowledgements}
% %% ***************************************************************************** %%

This work was funded by generous grants from the European H2020 Marie Sklodowska-Curie action (grant no. 101003403, \url{http://basira-lab.com/normnets/}) to I.R. and the Scientific and Technological Research Council of Turkey to I.R. under the TUBITAK 2232 Fellowship for Outstanding Researchers (no. 118C288, \url{http://basira-lab.com/reprime/}). However, all scientific contributions made in this project are owned and approved solely by the authors.

%%%%%%%%%%%%%%%%%%%%%%%%%%%%%%%%%%%%%%%%%%%%%%%%%%%%%%%%%%%%%%%%%%%%%%%%%%%%%%%%%%%%%%%%%%%%%%%%%%%%%%%%%%%%
\bibliography{Biblio3}
\bibliographystyle{splncs}
\end{document}